\documentclass[a4paper,12pt]{article}

\newcommand{\sect}[1]{\setcounter{equation}{0}\section{#1}}

\begin{document}
\topmargin 0pt \oddsidemargin 0mm

\renewcommand{\thefootnote}{\fnsymbol{footnote}}
\begin{titlepage}
\begin{flushright}
hep-th/0306140
\end{flushright}

\vspace{5mm}
\begin{center}
{\Large \bf Boosted Domain Wall and Charged Kaigorodov Space }
\vspace{12mm}

{\large
Rong-Gen Cai\footnote{Email address: cairg@itp.ac.cn}\\
\vspace{8mm}
{ \em Institute of Theoretical Physics, Chinese Academy of Sciences, \\
   P.O. Box 2735, Beijing 100080, China}}
\end{center}
\vspace{5mm} \centerline{{\bf{Abstract}}}
 \vspace{5mm}
 The Kaigorodov space is a homogeneous Einstein space and it describes a pp-wave
 propagating in anti-de Sitter space. It is conjectured in the literature that
 M-theory or string theory on the Kaigorodov space times a compact manifold is dual
 to a conformal field theory in an infinitely-boosted frame with constant momentum
 density. In this note we present a charged generalization of the Kaigorodov space by
 boosting a non-extremal charged domain wall to the ultrarelativity limit where the boost
 velocity approaches the speed of light. The finite boost of the domain wall solution
 gives the charged generalization of the Carter-Novotn\'y-Horsk\'y  metric. We study the
 thermodynamics associated with the charged Carter-Novotn\'y-Horsk\'y space and discuss its
 relation to that of the static black domain walls and its implications in the domain
 wall/QFT (quantum field theory) correspondence.

\end{titlepage}

\newpage
\renewcommand{\thefootnote}{\arabic{footnote}}
\setcounter{footnote}{0} \setcounter{page}{2}

%%=====================section 1====================
\sect{Introduction}
 In recent years there has been considerable interest in
 anti-de Sitter (AdS) space and asymptotically AdS space.  This
 should be mainly attributed to the celebrated AdS/CFT (conformal
 field theory) correspondence~\cite{AdS}, which says that there is a duality
 between M-theory or string theory on the AdS space times a compact space
 and a strongly coupling conformal field theory residing on the
 boundary of the AdS space.  In the spirit of the  AdS/CFT
 correspondence, it was convincingly argued by Witten~\cite{Witten} that the
 thermodynamics of black holes in AdS space can be identified with
 that of the dual CFT at high temperature.  Further it was proposed by
 de Boer, Verlinde and Verlinde \cite{Verlinde} that
 there is a correspondence between the classical evolution equations of bulk supergravity
 and the renormalization group equations of the dual boundary conformal field theory.

 The Kaigorodov space~\cite{Kaig} is an homogeneous Einstein space and
 it describes a pp-wave propagating
 in a four dimensional AdS space~\cite{Podo}. Its higher dimensional
 generalization was given in \cite{CLP}. In that reference,
 it was found that the Kaigorodov space naturally appears as the
 near-horizon geometry of non-dilatonic $p$-branes superimposed by
 a pp-wave propagating along a direction in the worldvolume of the
 $p$-branes. For example, five-, four- and seven-dimensional
 Kaigorodov spaces arise as the near-horizon geometries of D3-branes
 in type IIB string theory, M2-branes and M5-branes in M theory, respectively.
 Following the AdS/CFT correspondence, naturally it can be conjectured that M
 theory or string theory on the Kaigorodov space times a compact space is dual to
 a certain conformal field theory in an infinitely-boosted frame
 with  finite momentum density~\cite{CLP}(see also \cite{BCR}).
 Further, it was found that the
 Carter-Novotn\'y-Horsk\'y (CNH) metric~\cite{CNH}, which is a solution of
 the Einstein equations with a negative cosmological constant, occurs
 as the near-horizon geometry of non-extremal non-dilatonic
 $p$-branes superimposed by a pp-wave propagating along a
 direction of the worldvolume of the $p$-branes\footnote{In the case of non-extremal
 $p$-branes it seems more suitable to say that there is a finite momentum, instead of a pp-wave,
 imposed on the $p$-branes.  A pp-wave propagates in the speed of light, but the boost
 velocity is less than the speed of light in the case of non-extremal
 $p$-branes.}. Having considered possible application in the AdS/CFT correspondence, it
 is worthwhile to further discuss the Kaigorodov space and
 CNH space. On the other hand, they are of interest in their own
 right since they are exact solutions of clearly physical meanings to the Einstein equations
 with a negative cosmological constant.

 In this short note we present a charged generalizations of the
 Kaigorodov space and the CNH space by boosting a class of
 charged domain wall solution. They are exact solutions to the
 Einstein-Maxwell equations with a negative cosmological constant.
 The charged Kaigorodov space has an explanation as a pp-wave
 propagating in the domain wall background imposed a null gauge
 field  and the charged CNH space comes from the finite boost of the
 domain wall space. It is argued that gauged supergravity on the Kaigorodov
 space is dual to a certain strongly coupling CFT with R-charge in an infinitely
 boosted frame with finite momentum density. We will also discuss the thermodynamics
 of the charged CNH space, its relation to that of static domain wall and
 its implications in the generalized AdS/CFT correspondence.

%%=================section2===================
\sect{Charged Kaigorodov Space}
 Let us start with an $(n+2)$-dimensional ($n\ge 2)$ charged topological black hole
 solution~\cite{CaiSoh,Cham}
 \begin{eqnarray}
 \label{2eq1}
  && ds^2 =-f(r) dt^2 +f^{-1}(r)dr^2 +r^2 d\Sigma_{n,k}^2
  \nonumber
  \\
  && A= -\frac{1}{c}\frac{\tilde q}{r^{n-1}}dt, \ \ \
  c=\sqrt{\frac{2(n-1)}{n}},
  \end{eqnarray}
  where
  \begin{equation}
  f(r) = k -\frac{m}{r^{n-1}} +\frac{\tilde q^2}{r^{2(n-1)}}
  +\frac{r^2}{l^2}.
  \end{equation}
Here $m$ and $\tilde q$  are two integration constants,
$d\Sigma_{n,k}^2$ denotes the line element of an $n$-dimensional
Einstein space with constant curvature, $n(n+1)k$. Without loss of
 generality, one can take the value of $k$ as $\pm 1$ and $0$. It
is easy to check that the solution (\ref{2eq1}) is an exact
solution of the following Einstein-Maxwell equations with a
negative cosmological constant, $-n(n+1)/2l^2$,
\begin{eqnarray}
\label{2eq3}
 && R_{\mu\nu}-\frac{1}{2}g_{\mu\nu}R
-\frac{n(n+1)}{2l^2}g_{\mu\nu} = 2F_{\mu\lambda}F^{\lambda}_{\
\nu}-\frac{1}{2}g_{\mu\nu}F^2, \nonumber \\
&& \partial_{\mu}(\sqrt{-g}F^{\mu\nu})=0.
\end{eqnarray}
When $k=1$, the hypersurface $\Sigma_{n,1}$ could be an
$n$-dimensional sphere. In that case, the solution (\ref{2eq1})
represents an $(n+2)$-dimensional spherically symmetric charged
black hole in AdS space. When $k=-1$, $\Sigma_{n,-1}$ is an
$n$-dimensional  negative constant curvature space and it could be
a closed hypersurface with arbitrary high genus under appropriate
identification. In this note we are interested in the case of
$k=0$ with $\Sigma_{n,0}$ being an $n$-dimensional Euclidean space
\begin{equation}
\label{2eq4}
 dE_n^2\equiv d\Sigma^2_{n,0}=(dx_1^2 +dx_2^2+ \cdots
 +dx_n^2)/l^2,
 \end{equation}
 where we have rescaled the coordinates $x_i$: $x_i \to x_i/l$.
 Clearly, in this case, the solution describes a black domain wall
 spacetime\footnote{In \cite{CaiZh} such type of solutions is
 called black plane solutions.}. To see this, let us take the
 limit $r\to \infty$, with which one can find that the
 Poincar\'e symmetry is recovered on the worldvolume $(t,x_i)$ of the domain
 wall.  $r$ is the transverse coordinate of the domain
 wall.  To more clearly see this, let us rewrite the solution (\ref{2eq1}) in terms
 of ``isotropic" coordinates. Defining~\cite{Cai}
 \begin{equation}
 \label{2eq5}
 m = \mu +2q, \ \  \tilde {q}^2 =q (\mu +q), \ \
 r^{n-1} \to r^{n-1} +q,
 \end{equation}
 we can change (\ref{2eq1}) to the following form
 \begin{eqnarray}
  \label{2eq6}
 && ds^2 =\frac{r^2H^{2/(n-1)}}{l^2} \left [-\left (1-\frac{\mu l^2}{r^{n+1}}H^{-2n/(n-1)}\right)
     dt^2  + dE^2_n \right ]
 +H^{2/(n-1)}\tilde f^{-1}(r)dr^2, \nonumber \\
 && A =-\frac{1}{c}\frac{\tilde q}{r^{n-1}+q}dt,
 \end{eqnarray}
 where
 \begin{equation}
\tilde f(r) = -\frac{\mu}{r^{n-1}} +\frac{r^2}{l^2} H^{2n/(n-1)},
 \ \ \  H=1 +\frac{q}{r^{n-1}}.
 \end{equation}
 When the Schwarzschild parameter $\mu =0$ and keep $q$ as a constant, the solution is
supersymmetric, preserving 1/2 supersymmetries of gauged
supergravity. However, $r=0$ is a naked singularity in this case.
In these coordinates, the Poincar\'e symmetry on the worldvolume
is evident once one takes $\mu =0$. This makes it possible to
impose a pp-wave propagating on the domain wall~\cite{Garf}.

Now we make a boost transformation along  one of worldvolume
coordinates, say $x_1$, as follows,
\begin{equation}
\label{2eq8}
 t \to t \cosh\alpha - x_1 \sinh\alpha , \ \ \ x_1 \to
-t \sinh\alpha+ x_1 \cosh \alpha,
\end{equation}
where $\alpha$ is the boost parameter. The solution (\ref{2eq1})
is then changed to
\begin{eqnarray}
\label{2eq9}
 && ds^2=\frac{r^2}{l^2}\left (-dt^2
 +dx_1^2+\left(\frac{ml^2}{r^{n+1}}
 -\frac{\tilde q^2 l^2}{r^{2n}}\right) d(\cosh \alpha dt-\sinh\alpha dx_1)^2
 +dE^2_{n-1} \right) \nonumber \\
 &&~~~~~~+f^{-1}(r)dr^2, \nonumber \\
 && A =-\frac{1}{c}\frac{\tilde q}{r^{n-1}}(\cosh\alpha dt
   -\sinh \alpha dx_1),
 \end{eqnarray}
 where $dE^2_{n-1} =dx_2^2 +\cdots +dx_n^2$. Next we take the
 limit $\alpha \to \infty$. In that case, the transformation
 (\ref{2eq8}) is singular. To get a well-defined new solution to
 the Einstein-Maxwell equations with a negative cosmological
 constant, we take $m \to 0$ and $\tilde q \to 0$ and keep $m \cosh^2 \alpha = P$ and
 $\tilde q \cosh \alpha =P_e$ as two  constants while $ \alpha \to
 \infty$. With this approach  we obtain
 \begin{eqnarray}
 \label{2eq10}
 && ds^2 =\frac{r^2}{l^2} \left ( du dv  + \left(
 \frac{Pl^2}{r^{n+1}}-\frac{P^2_el^2}{r^{2n}} \right)du^2
  +dE^2_{n-1}\right ) + \frac{l^2}{r^2}dr^2, \nonumber \\
 && A = -\frac{1}{c}\frac{P_e}{r^{n-1}}du,
\end{eqnarray}
where $u,v = x_1 \mp t$ are two null coordinates. It is easy to
check that the new solution (\ref{2eq10}) satisfies the
Einstein-Maxwell equations (\ref{2eq3}) with a negative
cosmological constant. When $P_e=0$, the solution (\ref{2eq10})
reduces to the higher dimensional generalization of the Kaigorodov
space  obtained in~\cite{CLP}. Evidently the solution
(\ref{2eq10}) represents a pp-wave propagating along the direction
$v$. The constant $P$ has the physical interpretation as the
momentum density of the pp-wave. The term involving $P_e$ comes
from the contribution of the null Maxwell field. The solution
(\ref{2eq10}) is just the charged generalization of the Kaigorodov
space. As the Kaigorodov space~\cite{CLP}, we expect the charged
generalization (\ref{2eq10}) also preserves 1/4 supersymmetries of
gauged supergravity.  Note that the solution (\ref{2eq10}) is
asymptotically AdS. When $P=P_e=0$, the solution is just the AdS
space in the Poincar\'e coordinates, it can  be viewed as a
special domain wall space. As a result, the charged Kaigorodov
space can also be explained as a pp-wave propagating on the domain
wall with a null Maxwell field.

In the AdS/CFT correspondence, the bulk Maxwell field could be
dual to the gauge field of R-symmetry of boundary conformal field
theory and the bulk gauge charge could be identified to the
R-charge of boundary field theory~\cite{Cham,Gubs}. Following the
proposal that M-theory or string theory on the Kaigorodov space
times a compact space is dual to a certain strongly coupling
conformal field theory in an infinitely boosted frame with a
finite momentum density~\cite{CLP}, and the domain wall/QFT
(quantum field theory) correspondence which says that gauge
supergravity on a domain wall could be equivalent to a certain
quantum field theory living on the domain wall~\cite{dw/qft}, we
have the following generalization: gauge supergravity on the
charged Kaigorodov space (\ref{2eq10}) is dual to a conformal
field theory with  R-charge in an infinitely boosted frame with
finite momentum density.

%%==================section 3====================
\sect{Charged Carter-Novotn\'y-Horsk\'y space and associated
thermodynamics}

From the Lorentz transformation (\ref{2eq8}) we know the boost
velocity $v$ is
\begin{equation}
\label{3eq1}
 v =\tanh \alpha.
\end{equation}
When $\alpha \to \infty$, the boost velocity approaches to the
speed of light. Therefore the charged Kaigorodov metric
(\ref{2eq10}) is obtained by boosting the charged domain wall
solution to the ultrarelativity limit $v=1$. Note that in this
limit the Lorentz transformation (\ref{2eq8}) is singular.
However, the resulting new solution (\ref{2eq10}) is well-defined.

Now we discuss the case for a finite boost, namely the case when
the boost parameter $\alpha$ is a finite constant. In this case we
notice that when $\tilde q =0$, the solution (\ref{2eq9}) gives us
the Carter-Novotn\'y-Horsk\'y (CNH) metric~\cite{CNH,CLP},
describing a non-extremal black domain wall with a pp-wave
propagating on the worldvolume, or say, a boosted black domain
wall with a finite momentum density. Therefore our solution
(\ref{2eq9}) gives a charged generalization of the CNH space.
Clearly the solution (\ref{2eq9}) is locally equivalent to the
static solution (\ref{2eq1}) since the former comes from the
latter via the coordinate transformation (\ref{2eq8}). However,
this transformation is valid only if the coordinate $x_1$ is not
periodic~\cite{CLP}. Therefore if the coordinate $x_1$ is
periodic, the solution (\ref{2eq9}) is not equivalent to the
static one (\ref{2eq1}) globally.

In \cite{Witten} it is argued that the thermodynamics of black
holes in AdS space can be identified with that of the dual CFT
residing on the boundary of the AdS space. To check the
gravity/field theory duality that gauged supergravity on the
charged Kaigorodov space (\ref{2eq10}) is dual to a dual CFT in an
infinitely boosted frame with a finite momentum density, we now
show that the thermodynamics associated with the static solution
(\ref{2eq1}) and boosted solution (\ref{2eq9}) can indeed be
connected via a Lorentz transformation.

For the static black domain wall solution (\ref{2eq1}), the mass
density  and the electric charge density on the domain wall are
\begin{equation}
\label{3eq2}
 \textbf{m} = \frac{M}{L_1V_{n-1}}=\frac{n m}{16\pi G l^n},\ ~~~~~~~ \
  \textbf{q} = \frac{Q}{L_1V_{n-1}}=\frac{\sqrt{2n(n-1)}}{8\pi G l^n}\tilde q,
\end{equation}
and the chemical potential conjugate to the electric charge is
\begin{equation}
\Phi= \frac{\tilde q}{cr_+^{n-1}}.
\end{equation}
Here $L_1$ is the length of the coordinate $x_1$ in (\ref{2eq4}),
$V_{n-1}$ denotes the volume of the hypersurface spanned by
coordinates $x_2$, $\cdots$, $x_n$, and $r_+$ stands for the
horizon location of  the black domain wall (\ref{2eq1}), the large
root of the equation $f(r_+)=0$. Note that the coordinates $x_i$
are dimensionless so that here  the length $L_1$ and the volume
$V_{n-1}$ are also dimensionless. The Hawking temperature $T$ and
entropy density of the static solution are found to be
\begin{equation}
\label{3eq4}
 T=\frac{1}{4\pi r_+}\left ((n+1)\frac{r_+^2}{l^2}
  -(n-1)\frac{\tilde q^2}{r_+^{2(n-1)}} \right),
\ ~~~~~\ \textbf{s}=\frac{S}{L_1V_{n-1}}=\frac{r_+^n}{4Gl^n}.
\end{equation}
These thermodynamic quantities obey the first law of black hole
thermodynamics,
\begin{equation}
\label{3eq5}
 d\textbf{m}= Td\textbf{s} +\Phi d\textbf{q}.
 \end{equation}

 On the other hand, for the boosted solution, the horizon location $r_+$ is still
 determined by the equation $f(r_+)=0$. From the solution (\ref{2eq9}), we
 obtain the mass density and electric charge density
 \begin{equation}
 \label{3eq6}
 \textbf{m}'=\frac{M'}{L_1'V_{n-1}}=\frac{nm}{16\pi G l^n}\cosh^2 \alpha,  \ ~~~~~ \
 \textbf{q}'=\frac{Q'}{L_1'V_{n-1}}=\frac{\sqrt{2n(n-1)}}{8\pi G l^n}\tilde
 q  \cosh \alpha,
 \end{equation}
 and the chemical potential associated with the charge is
 \begin{equation}
 \label{3eq7}
 \Phi '= \frac{\tilde q}{cr_+^{n-1}}\cosh \alpha.
 \end{equation}
 Besides, the solution (\ref{2eq9}) has the momentum density,
 which can be read off directly from the solution,
 \begin{equation}
 \label{3eq8}
 \textbf{p}'=\frac{P_v'}{L_1'V_{n-1}}=\frac{nm}{16\pi G l^n}\sinh \alpha\cosh
 \alpha.
 \end{equation}
 The conjugate quantity to the momentum is the boost velocity, $ v=\tanh\alpha$.
 Here $L_1'$ denotes the length of the coordinate $x_1$ in the
 solution (\ref{2eq9}) and the volume of the hypersurface
 described by the line element $dE^2_{n-1}$ has been assumed the same as the one
 for the static solution (\ref{2eq1}). In addition, we find that
 the Hawking temperature and entropy density are changed to
 \begin{eqnarray}
 \label{3eq9}
&& T'=\frac{1}{4\pi r_+\cosh \alpha}\left ((n+1)\frac{r_+^2}{l^2}
  -(n-1)\frac{\tilde q^2}{r_+^{2(n-1)}} \right)=T/\cosh \alpha,
\nonumber \\
&& \textbf{s}'=\frac{S'}{L_1'V_{n-1}}=\frac{r_+^n}{4Gl^n}\cosh
\alpha.
\end{eqnarray}
Contrast to the first law (\ref{3eq5}), due to the appearance of
the momentum density, one might think that the boosted black
domain wall solution (\ref{2eq9}) has the following first law
\begin{equation}
\label{3eq10}
 d\textbf{m}'=T'd\textbf{s}'+\Phi'd\textbf{q}' +v d\textbf{p}'.
\end{equation}
Substituting those quantities into (\ref{3eq10}), however, it is
found that this expression does not hold at all.  Have a look
again at the solution (\ref{2eq9}), we find that due to the boost,
there is an electric  current along the direction $x_1$. The
current density is found to be
\begin{equation}
\label{3eq11}
\textbf{j}'=\frac{J'}{L_1'V_{n-1}}=\frac{\sqrt{2n(n-1)}}{8\pi G
l^n}\tilde q  \sinh \alpha = v \textbf{q}' ,
\end{equation}
and the conjugate chemical potential is
\begin{equation}
\label{3eq12}
 \Psi' =-\frac{\tilde q}{cr_+^{n-1}}\sinh \alpha.
\end{equation}
With this component, we obtain the correct first law of
thermodynamics associated with the boosted charged domain wall
\begin{equation}
\label{3eq13}
 d\textbf{m}'=T'd\textbf{s}'+\Phi'd\textbf{q}' +
\Psi'd \textbf{j}' + v d\textbf{p}'.
\end{equation}
Now it is an easy job to check that those thermodynamic quantities
indeed satisfy the first law (\ref{3eq13}).

According to the generalized AdS/CFT correspondence, the
thermodynamics of the static charged domain wall (\ref{2eq1}) can
be identified with that of the dual CFT with a R-charge in a
static frame, while the thermodynamics associated with the boosted
domain wall (\ref{2eq9}) can be mapped to that of the same CFT in
a frame with a finite boost with velocity $v =\tanh \alpha$. To
check this correspondence, it is interesting to compare the
thermodynamics of the boosted and static domain walls. We know
that entropy accounts for the number of independent quantum states
of a system. Therefore, the entropy should remain unchanged within
the Lorentz boost transformation. We see from (\ref{3eq4}) and
(\ref{3eq9}) that indeed the total entropy remain invariant $S=S'$
once taking into account the relation
\begin{equation}
\label{3eq14}
 L_1'= L_1/\cosh \alpha,
 \end{equation}
 which is just the Lorentz relation between these two different frames. The
 electric charge $Q$ (corresponding to the R-charge in the
 dual CFTs) is a conserved quantity, which should be also
 invariant within the Lorentz transformation. From (\ref{3eq2})
 and (\ref{3eq6}) we have indeed $Q=Q'$ once considering the
 relation (\ref{3eq14}). Further because the Hawking temperature
 of event horizon is related to what coordinates one used are,
one can see from (\ref{3eq4}) and (\ref{3eq9}) they are different.
However, they are related to each other via the relation
$T'=T/\cosh\alpha$, which is just the manifest of the Lorentz
transformation. In addition, in the static coordinates the
thermodynamic system has energy $M$ and vanishing momentum, while
in the boosted frame the energy is changed to  $M'=M \cosh
\alpha$, and the momentum is non-vanishing, $P'_v=v
M/\sqrt{1-v^2}$. These are just the Lorentz transformation
relation for the four-momentum. In a word, we verify that those
thermodynamic quantities of CFTs dual to the boosted and static
black domain walls are indeed related via the Lorentz
transformation. Therefore, this result provides support of the
conjecture that gauged supergravity on the Kaigorodov space is
dual to a boundary CFT in an infinitely-boosted frame with a
finite momentum density.

%%====================section 4====================
\sect{Conclusion}
 In summary we have presented a charged generalization of the
 Kaigorodov metric by boosting a charged black domain wall
 solution to the ultrarelativity limit where the boost velocity
 approaches to the speed of light. Although the limit is singular,
 the resulting new solution, the charged Kaigorodov space, is
 well-defined. The solution is an exact solution to the
 Einstein-Maxwell equations with a negative cosmological
 constant and describes a pp-wave propagating on an AdS domain
 wall with a null Maxwell field.

 It is conjectured that gauge supergravity on the charged
 Kaigorodov space is dual to a strongly coupling CFT with
 R-charge in an infinitely boosted frame with a  finite momentum
 density. To show this, we have discussed the thermodynamics of
 the charged Carter-Novotn\'y-Horsk\'y (CNH) solution, which is a
 result of a finite boost for the charged black domain wall
 metric, and have compared the thermodynamics of booted and static
 black domain walls. According to the generalized AdS/CFT
 correspondence, the thermodynamics of the charged black domain
 wall can be identified with that of dual CFT with R-charge. We
 have shown that the thermodynamic quantities associated with
 the charged CNH solution indeed can be related to those of
 static domain wall solution through a Lorentz transformation.
 Note that both the charged Kaigorodov space and charged CHN space
 discussed in this note are asymptotically AdS. It is therefore of
 interest to further discuss this correspondence to the case of
 asymptotically non-AdS domain wall spaces in the sense of the
 domain wall/QFT correspondence~\cite{dw/qft}.

%%========================references ==============
\section*{Acknowledgment}

This work was supported in part by a grant from Chinese Academy of
Sciences, a grant from the Ministry of Education of China and by
the Ministry of Science and Technology of China under grant No.
TG1999075401.

\end{document}